\documentclass[usegraphicx,useAMS,usenatbib]{mn2e}

\title[3D Spectroscopy of Local Luminous Compact Blue Galaxies: Kinematics of NGC 7673]
{3D Spectroscopy of Local Luminous Compact Blue Galaxies: Kinematics of NGC 7673}
\author[J. P\'erez-Gallego et al.]
{J. P\'erez-Gallego,$^{1}$\thanks{E-mail: jgallego@astro.ufl.edu} 
R. Guzm\'an,$^{1}$
A. Castillo-Morales,$^{2}$
F. J. Castander,$^{3}$
J. Gallego,$^{2}$
\newauthor C. A. Garland,$^{4}$
N. Gruel,$^{1}$
D. J. Pisano,$^{5}$
S. F. S\'anchez,$^{6}$ and
J. Zamorano$^{2}$
\\
$^{1}$Department of Astronomy, University of Florida, 211 Bryant Space Science Center, Gainesville, FL 32611-2055, USA\\
$^{2}$Departamento de Astrof\'isica, Universidad Complutense de Madrid, 28040 Madrid, Spain\\
$^{3}$Institut de Ci\`encies de l'Espai (ICE/CSIC), Campus UAB, 08193 Bellaterra, Barcelona, Spain\\
$^{4}$Department of Natural Sciences, Black Science Center, Castleton State College, Castleton, VT 05735, USA\\
$^{5}$Department of Physics, Hodges Hall, West Virginia University (WVU/NRAO), Morgantown, WV 26506-6315, USA\\ 
$^{6}$Centro Astron\'omico Hispano Alem\'an, AIE (CSIC-MPG), E-04004 Almeria, Spain}
\begin{document}

\date{Accepted 2009 November 4.  Received 2009 October 26; in original form 2009 May 8}

\pagerange{\pageref{firstpage}--\pageref{lastpage}} \pubyear{2009}

\maketitle

\label{firstpage}

\begin{abstract}
The kinematic properties of the ionized gas of local Luminous Compact Blue Galaxy (LCBG) NGC 7673 are presented using three dimensional data taken with the PPAK integral field unit at the 3.5-m telescope in the Centro Astron\'omico Hispano Alem\'an. Our data reveal an asymmetric rotating velocity field with a peak to peak difference of 60 km s$^{-1}$. The kinematic centre is found to be at the position of a central velocity width maximum ($\sigma=54\pm1$ km s$^{-1}$), which is consistent with the position of the luminosity-weighted centroid of the entire galaxy. The position angle of the minor rotation axis is 168$^{\circ}$ as measured from the orientation of the velocity field contours. At least two decoupled kinematic components are found. The first one is compact and coincides with the position of the second most active star formation region (clump B). The second one is extended and does not have a clear optical counterpart. No evidence of active galactic nuclei activity or supernovae galactic winds powering any of these two components has been found. Our data, however, show evidence in support of a previously proposed minor merger scenario in which a dwarf galaxy, tentatively identified with clump B, is falling into NGC 7673. and triggers the starburst. Finally, it is shown that the dynamical mass of this galaxy may be severely underestimated when using the derived rotation curve or the integrated velocity width, under the assumption of virialization.
\end{abstract}

\begin{keywords}
galaxies: starburst -- galaxies: individual:NGC 7673.
\end{keywords}

\section{Introduction}

Galaxies in which star formation and associated phenomena dominate the total energetics are known as starburst galaxies \citep{weedman83}. These objects have a large star formation rate per unit area compared to normal galaxies, and the time it would take to produce the current stellar mass at the current star formation rate is much less than the age of the Universe. These equivalent definitions \citep{kennicutt98} cover galaxies with a wide variety of properties at different redshifts. Distant Lyman Break Galaxies (LBGs) \citep{steidel96,lowenthal97}, and closer Blue Compact Dwarfs (BCDs) \citep{cairos01} and Luminous Compact Blue Galaxies (LCBGs) \citep{werk04} fall under this category. This denotes their cosmological relevance and turns local starburst galaxies into unique laboratories to study the complex ecosystem of the star formation process throughout time when they can be properly and equally selected at different epochs of the Universe.

LCBGs are small starburst systems that dominate the number density of optically selected galaxies at $z\sim1$ \citep{ferguson04}. The most comprehensive study of LCBGs at intermediate redshift to date is that of \citet{phillips97} and \citet{guzman97}, who concluded that the LCBG class is a populated mixture of starbursts. About 60\% of galaxies in their sample are classified as ``H {\small{II}}-like'' since they are similar to today's population of luminous, young, star-forming H {\small{II}} galaxies. The remaining 40\% are classified as ``SB disk-like'' since they form a more heterogeneous class of evolved starbursts similar to local starburst disk nuclei and giant irregular galaxies. 

From an observational point of view, LCBGs are, as described by \citet{werk04}, objects with (i) absolute blue magnitude ($M_B$) brighter than -18.5; (ii) effective surface brightness ($SB_e$) brighter than 21 $B$-mag~arcsec$^{-2}$; and (iii)($B-V$) color bluer than 0.6. In this observational parameter space defined by $M_B$, $SB_{e}(B)$, and ($B-V$), LBGs and the intermediate redshift LCBGs share the same optical properties. One can argue that the sharp borders used to classify them are artificial, but they serve the purpose of identifying similar objects over a very large range of redshift, from $z\sim0$ to $z\sim2$. For instance, if one considers the limiting magnitude of deep wide-field Hubble Space Telescope imaging ($I_{AB}=26$ for extended objects with $SB_e(I_{AB})=24.5$~mag~arcsec$^{-2}$ according to \citet{scoville07}), these limits correspond to rest-frame $M_B\sim-18.5$ and $SB_e(B)\sim21$~mag~arcsec$^{-2}$. These selection criteria identify objects not only with similar observational properties but also with similar physical properties. The galaxies in this morphologically heterogeneous starburst population form stars at around 10 \--- 20~$M_{\odot}$~yr$^{-1}$, show velocity widths of 30 \--- 120~km~s$^{-1}$, are as compact as $R_e\sim2$ \--- 5~kpc, and have metallicitites lower than solar. 

In the last decade, various observational studies have highlighted the key role that LCBGs play in galaxy evolution over cosmological time scales. We know now that LCBGs contribute the most to the evolution of the blue $L^{\star}$ galaxies in the last $\sim8$~Gyrs \citep{lilly98,mallen99,melbourne07}. LCBGs are as well a major contributor to the observed increase in the star formation rate density of the universe from $z=0$ to $z\sim1$ \citep{guzman97,noeske07,vergani08}. And, finally, and more interesting from a cosmological point of view, LCBGs may be lower mass counterparts of the star-forming galaxy population at $z\sim3$, including LBGs \citep{steidel03}.

In addition, LCBGs might hold the key to understanding the physical mechanisms that link the red and blue galaxy sequences. This division is one important observational result derived from color-magnitude diagrams \citep{cattaneo06}, but also seen in the luminosity function \citep{baldry04,bell03}, the joint color and stellar mass distribution \citep{baldry04,balogh04,hogg04}, the stellar age and the star formation rate (SFR) \citep{kauffmann03}, the gas-to-stellar mass ratio \citep{kannappan04}, and their local environment \citep{blanton06}. Current models explain both a truncation in the stellar mass of the blue sequence at $M \sim3\times10^{10}$~$M_{\odot}$, and the break out of the red sequence at z $\sim 1.5$ \citep{cattaneo06}. Galaxies grow along the blue sequence until gas accretion is halted by the critical shock-heating mass of the host halo. They then redden and fade in luminosity, moving up to the red sequence, whose bright end is reached either by dissipationless or gas-rich mergers. The majority of $L^{\star}$ galaxies at $z>1$ are intrinsically bright, small and blue, and lie close to the critical stellar mass of the blue sequence and at the bright end of the ubiquitous intermediate-$z$ blue galaxy population that dominates the number counts in deep optical surveys. Therefore, LCBGs, which populate the high mass end of the blue sequence, are key to explaining any scenario that deals with the division of galaxies into a blue and a red sequence.

Furthermore, the lack of LCBGs in the local universe \citep{werk04} raises the issue of how such an ubiquitous population has evolved over the last 9 \--- 10~Gyr. Some authors suggest LCBGs may be the progenitors of today's spheroidal galaxies and/or the spheroidal component of today's disk galaxies; both evolutionary scenarios are in agreement with the \citet{cattaneo06} model. The former are low-mass ($M<10^{10}$~$M_{\odot}$) spheroidal galaxies or dwarf elliptical galaxies whose properties according to evolutionary models would be matched by a typical LCBG after a 4 \--- 6~Gyr fading process \citep{koo94,guzman98,noeske06}. The latter are present day small spirals, more massive ($M\sim10^{10}$~$M_{\odot}$) than inferred from virial masses, whose emission is mostly due to a vigorous central burst \citep{phillips97,hammer01,puech06}. 

A key ingredient in the discussion of whether LCBGs evolve one way or another is their mass. A reliable determination of their masses is necessary to properly place LCBGs within one evolution scenario or the other, or to understand what makes them evolve one way or the other. Masses of LCBGs can be derived from their rotation curves and velocity widths, nevertheless one needs to be careful with those, since most of these objects' kinematics might not be coupled to their masses due to supernova galactic winds, and minor and major mergers. While a thorough study of the kinematics of each galaxy solves this issue at lower redshifts where we might be able to spatially resolve any kinematic decoupled component, at higher redshifts we need to be careful when coupling these objects' kinematics to their masses.      

Interest in LCBGs has multiplied following the initial observational results and theoretical models highlighted above, but their relation to today's galaxy population still remains unknown. In order to understand the nature of LCBGs and its role in galaxy evolution we have selected a representative sample of 21 LCBGs within 100~Mpc from the Sloan Digital Sky Survey (SDSS) \citep{adelman06}, Universidad Complutense de Madrid (UCM) \citep{zamorano94}, and Markarian \citep{markarian89} catalogs that best resemble the properties of the distant LCBGs, ensuring that this sample, although small, is representative of the LCBG  population as a class by covering the whole range in luminosity, color, surface brightness, and environment. We are carrying out a multiwavelength study of these nearby LCBGs, including the optical, millimeter, and centimeter ranges. In this paper we focus on the optical three dimensional (3D) spectroscopy of a single LCBG, NGC 7573. \citet{garland07} present our initial H{\small{I}} and CO results. The optical is the best understood spectral range in nearby galaxies, and systematic studies at intermediate- and high-$z$ with the new generation of near-infrared multi-object spectrographs and integral field units in 10-m class telescopes have already started \citep{forster06,puech06}. The derived 3D maps allow us to test model predictions on the origin and evolution of these massive starbursts. Insights about the role of mergers and supernova galactic winds can be obtained  by, for instance, identifying decoupled kinematic components within the velocity fields of these objects and cross-correlating these kinematic components with morphological information. Furthermore, clues about the trigger mechanism for the current burst in such galaxies may be found.

\begin{figure}
\begin{center}
\includegraphics[width=8cm]{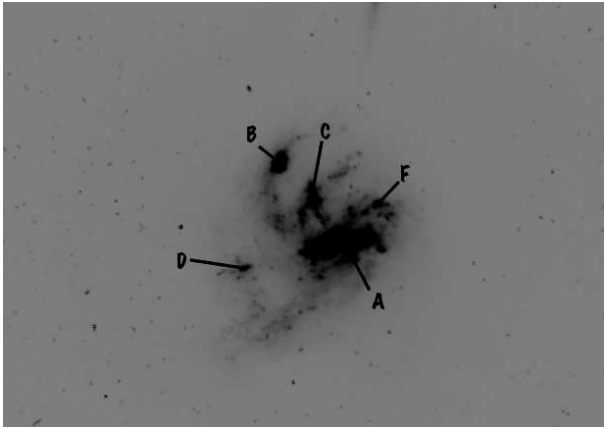}
\caption{F555W WFPC2 image of NGC 7673 labeled with the star-forming clumps identified by \citet{duflot82}. North is to the top and East is to the left. \label{fig:regions}}
\end{center}
\end{figure}  

NGC 7673 (also known as MRK 325 and UCM 2325+2318) is a nearby ($z = 0.011368$) starburst galaxy that has been widely studied in the past (e.g. Duflot-Augarde \& Alloin 1982; Homeier \& Gallagher 1999, hereafter HG99; Homeier et al. 2002; Pasquali \& Castangia 2008)
This irregular starburst galaxy (see Figure~\ref{fig:regions}) shows a clumpy structure with bright knots of star formation embedded in a diffuse halo that can be seen in the optical spectral range 
(HG99, P\'erez-Gonz\'alez et al. 2003). \citet{duflot82} found 6 different star-forming clumps (see Figure~\ref{fig:regions}), some of which are referred to in this paper. From PPAK H${\alpha}$ fluxes \citet{castillo09} derived a star formation rate of at least 6.4 $M_{\odot}$ yr$^{-1}$. Furthermore, its small size, high surface brightness, strong emission lines and blue colors make of NGC 7673 a prototypical LCBG (see Table~\ref{table:NGC7673properties}).

H {\small{I}} 21 cm line mapping of NGC 7673 and its environment \citep{nordgren97}, which includes neighboring galaxy NGC 7677, roughly 7' to the southeast (3554 vs. 3405~km~s$^{-1}$) (HG99) shows a small outer irregularity that points to the latter. Nevertheless, a present major merger scenario has been rejected because of a remarkably constant velocity across the galaxy \citep{duflot82}. HG99 used data taken with DENSEPAK to study the kinematics of a portion of the galaxy. They were limited by the pointing of the telescope and both the field of view (30 $\times$ 45~arcsec$^{2}$) and the spatial resolution of the instrument. Their spatial resolution sampling of two thirds of the galaxy, as seen in optical images, is highly improved by our sampling. Nevertheless, their spectral resolution (FWHM $=32$~km~s$^{-1}$) allowed them to find that a two component model, one narrow (FWHM $\sim55$~km~s$^{-1}$) and one broad (FWHM $\sim150$~km~s$^{-1}$) fit the observed spectra. According to their analysis three are the possible explanations for the presence of such a broad component: (i) a consequence of integrating over many ionized structures at different velocities; (ii) hot, turbulent gas confined to large cavities carved out by massive stars; and (iii) a starburst-powered galactic wind or similar break-out phenomenon. Furthermore, they exclude a present, but not past, interaction between NGC 7673 and NGC 7677, indicating a minor merger as the trigger mechanism for the major starburst. 

In this paper we focus on the optical kinematic properties of NGC 7673 and compare the kinematical properties of the ionized gas to those of the neutral atomic gas \citep{garland07}. Flux related properties, such as SFRs, metallicities, and extinction, are discussed in \citet[hereafter CM09]{castillo09}.

Throughout this paper we adopt the concordance cosmology, i.e., a flat universe with $\Omega=0.7$, $\Lambda=0.3$, and $h=0.7$. The paper is structured as follows. In Section 2 we describe our sample selection, observations and data reduction. In Section 3 we describe our measurements and results. The discussion is carried on in Section 4 and finally in Section 5 we state our conclusions and future plans.    

\section{Sample Selection, Observations, and Data Reduction}

\subsection{Sample Selection}\label{sample}

Our sample of LCBGs in the local universe was chosen from the the Data Release 4 (DR4) of the SDSS \citep{adelman06}, the UCM Survey \citep{zamorano94}, and the Markarian catalog \citep{markarian89}. The $B$ magnitudes ($M_B$), surface brightnesses ($SB_{e}$), colors ($B-V$), and sizes ($r_{e}$), of the candidate objects in these catalogs were taken into account to select them as described in the Introduction, i.e., $M_B<-18.5$; $SB_e>21$ $B$-mag~arcsec$^{-2}$; and($B-V$)$<0.6$. Furthermore, objects with sizes larger than 4 arcsec and closer than 100 Mpc were given priority to take full advantage of the instrument used. This yielded a sample of 22 nearby LCBGs. 

\begin{table*}
\caption{NGC 7673 Observational Properties}	
\begin{tabular}{lccccc}
\hline
{Name} & {$z$$^{a}$}  & {$M$($B$)$^{b}$} & {$SB_{e}$$^{b}$}  & {$B-V$$^{b}$} & {$R_e$$^{b}$} \\
{} & {}  & {(mag)} & {(mag arcsec$^2$)}  & {(mag)} & {(kpc)} \\	
\hline
NGC 7673   	& 0.011368$^{c}$ 		& $-20.50$ 		& 19.40  			& 0.30	 		& 1.9 \\
LCBGs$^{d}$	& $0.019\pm0.002$		& $-19.70\pm0.20$	& $20.2\pm0.5$ 	& $0.40\pm0.02$	& $2.2\pm0.3$ \\
\hline
\multicolumn{6}{l}{$^{a}$ \citet{huchra99}; $^{b}$ \citet{pisano01}; $^{c }$ This gives a scale of 0.230 kpc arcsec$^{-1}$}\\
\multicolumn{6}{l}{$^{d }$ Average and standard deviation from the local LCBGs in our sample}\\
\label{table:NGC7673properties}
\end{tabular}
\end{table*}

\begin{table*}
\caption{NGC 7673 Observational Log}
\begin{tabular}{lcccc}
\hline	
{Configuration} & {Dithering} & {Night}  & {Exposure Time} & {Offset}\\
{}  & {} & {}  & {(s)} & {(arcsec)}\\
\hline	
V300$^{b}$ 				& 1 & 10 August 2005 	& 3 $\times$ 330 		& (0.00, 0.00)$^{a}$	 \\
V300			 		& 2 &10 August 2005  	& 3 $\times$ 330 		& (1.56, 0.78)					 \\
V300			 		& 3 &10 August 2005 	& 3 $\times$ 330 		& (1.56, $-$0.78)				 \\
V1200$^{c}$		 		& 1 &11 August 2005 	& 3 $\times$ 900 		& (0.00, 0.00)$^{a}$	 \\
V1200			 		& 2 &11 August 2005 	& 3 $\times$ 900 		& (1.56, 0.78)					 \\
V1200			 		& 3 &11 August 2005 	& 3 $\times$ 900 		& (1.56, $-$0.78)	 			 \\
\hline
\multicolumn{5}{l}{$^{a}$The original pointing of the telescope is 23:27:41.5 +23:35:20.0}\\
\multicolumn{5}{l}{$^{b}$V300 centered at 5361 {\AA} covering the range between 3591 and 6996 {\AA}}\\
\multicolumn{5}{l}{$^{c}$V1200 centered at 5040 {\AA} covering the range between 4669 and 5395 {\AA}}\\
\label{table:NGC7673observations}	
\end{tabular}
\end{table*}

\subsection{Observations}\label{observations}

Objects from our sample were observed using PPAK \citep{kelz06} at the 3.5-m telescope in the Centro Astron\'omico Hispano Alem\'an\footnote{Based on observations collected at the German-Spanish Astronomical Center, Calar Alto, jointly operated by the Max-Planck-Institut f\"ur Astronomie Heidelberg and the Instituto de Astrof'sica de Andaluc\'ia (IAA/CSIC).} (CAHA). PPAK is an integral field unit (IFU) consisting of 331 science fibers with a diameter of 2.7 arcsec each, covering an hexagonal field of view (FOV) of 74 x 65~arcsec$^2$. Furthermore, there are 15 calibration fibers, and 36 sky fibers located 80 arcsec away from the centre of the hexagonal FOV. 

PPAK observations of NGC 7673 were made during the nights of 2005 August 10 and 11 using two different setups (see Table~\ref{table:NGC7673observations}). First, a 300~lines~mm$^{-1}$ grating (V300) centered at 5316~{\AA} was used. This low resolution configuration provided a spectral resolution of 10.7~{\AA} FWHM ($\sigma\sim255$~km~s$^{-1}$ at 5316~{\AA}) covering the spectral range from 3600 to 7000~{\AA}, including H$\beta$ and H$\alpha$. Three different dithering positions were observed to fill the gaps between each fiber and its next neighbors. Three exposures 330 s long were taken for a total exposure time of 990 s per dithering position. The offsets applied between the three different dithering positions allowed us to fully cover the field of view of the galaxy and overcome the presence of holes between fibers in the detector (see Table~\ref{table:NGC7673observations}).

Second, a 1200~lines~mm$^{-1}$ grating (V1200) centered at 5040~{\AA} was used. This high resolution configuration provided a nominal spectral resolution of 2.78~{\AA} FWHM ($\sigma\sim70$~km~s$^{-1}$ at 5040 {\AA}), covering the spectral range from 4900 to 5400~{\AA}, including H$\beta$ and [OIII]$\lambda$5007. Again, three different dithering positions were observed. Three exposures, each one 900 s long, were taken for a total exposure time of 2700 s per dithering position.

\subsection{Data Reduction}\label{reduction}

The data were reduced using R3D and E3D \citep{sanchez06}, IRAF\footnote{IRAF is distributed by the National Optical Astronomy Observatory, which is operated by the Association of Universities for Research in Astronomy (AURA) under cooperative agreement with the National Science Foundation.} and our own custom software. All the images were bias-subtracted, flat-fielded, and cosmic-ray cleaned. The 331 two-dimensional spectra per dithering position were then properly extracted, distortion-corrected, wavelength-calibrated, sky-subtracted, and flux-calibrated. Corrections were also applied to minimize the differences between fiber-to-fiber transmission. As for differential atmospheric refraction corrections, those were not applied for not being relevant (up to 0.2 and 0.4 arcsec in the x and y respectively). 

The V300 wavelength calibration was performed using a He lamp with eleven lines within the considered spectral range. The $rms$ of the best fit polynomial ($n=3$) was 0.12~{\AA}. A further analysis to evaluate the accuracy of our calibration including sky lines reveals a final uncertainty in our calibration of 0.2~{\AA} (10~km~s$^{-1}$ at 6000~{\AA}). For this analysis five high signal-to-noise (S/N) sky lines from the wave-calibrated spectra covering the entire V300 spectral range were fit by single Gaussian models for each fiber. The centroids of these lines were then compared to those of the CAHA sky atlas \citep{sanchez07}. The uncertainty of our measurements was given by the standard deviation of the resulting residuals, which is consistent within the entire V300 spectral range (i.e. for each sky line independently).

On the other hand, the V1200 wavelength calibration was performed using both He and Cs lamps because the He lamp lacked emission lines bluer than 5015~{\AA}. The Cs lamp was observed separately and both lamps were added to increase the spectral coverage. Furthermore, weak contamination lines from the Ar lamp used to illuminate the calibration fibers were also used for this purpose. The $rms$ of the best fit polynomial ($n=3$) translated into an uncertainty of at least 0.25 {\AA} (15~km~s$^{-1}$). The lack of sky lines did not allow us to carry on an analysis similar to the one carried out for the V300. While this is critical in the case we are interested in measuring the position of the centroid of the emission lines, it is not if we are only interested in measuring the velocity widths. Thus, our data can be wavelength-calibrated using the rest-frame positions of H$\beta$, [OIII]$\lambda$4959, and [OIII]$\lambda$5007, to measure velocity widths with the spectral resolution of the detector. For that, we wavelength-calibrated each of the fibers independently. When this is done the $rms$ for this particular spectral range is 0.13~{\AA} (7~km~s$^{-1}$).  

\section{Measurements and Results}

\subsection{Data Measurements}\label{measurements}

Emission lines were fit by single Gaussian profiles both in the low (V300, H$\alpha$)  and high resolution (V1200, [OIII]$\lambda$5007) configurations. A minimum S/N of 13 in the flux detection was required for the measurements to be considered. Below this the uncertainty associated with low S/N dominates over the wavelength calibration uncertainty as the main source of error, according to our simulations. 

In the outer areas of the galaxy, we spatially binned the V300 data by co-adding fibers to achieve a minimum S/N of 13. This way, we were able to obtain eight new measurements for dithering position one and two, and six for dithering position three. Furthermore, fibers from different dithering positions were co-added to obtain four new measurements. Each of these measurements were linked to the average position of the fibers co-added. These measurements make the final map extend towards the lower surface brightness outskirts of the galaxy sampling an area about 10\% bigger than before co-adding fibers.

The velocity map of the galaxy was derived from the H$\alpha$ emission lines for each dithering position with the low resolution spectra. The data were interpolated down to a spatial resolution of 1 arcsec pixel$^{-1}$ yielding a 60 $\times$ 60~pixel square grid. Each of the original fibers (i.e., original spatial resolution element) is therefore sampled by approximately 3 $\times$ 3~pixel$^2$ (i.e., $\sim2.7\times 2.7$~arcsec$^2$). This keeps us from oversampling the data, which may result in the appearance of artifacts \citep{sanchez06}. The three maps were then registered and averaged pixel by pixel. Thus, both the spatial resolution and the error associated with each measurement improved by a factor of $\sqrt{3}$ (6~km~s$^{-1}$ for the V300 and 4~km~s$^{-1}$ for the V1200). 

The velocity width map of the galaxy was derived from the [OIII]$\lambda$5007 emission lines for each dithering position with the high resolution spectra. This line was selected as the one with highest S/N within the V1200 spectral range. The instrumental broadening as measured in sky lines for each fiber was properly subtracted. The final map was produced as explained above with the difference that no binning was pursued in this case since it would artificially broaden the emission lines. Furthermore, as in HG99, an attempt to find multiple spectral components was made but the result was not conclusive due to insufficient spectral resolution. 

\subsection{Velocity Map}\label{kinvel} 

\begin{figure}
\begin{center}
\includegraphics[width=8cm]{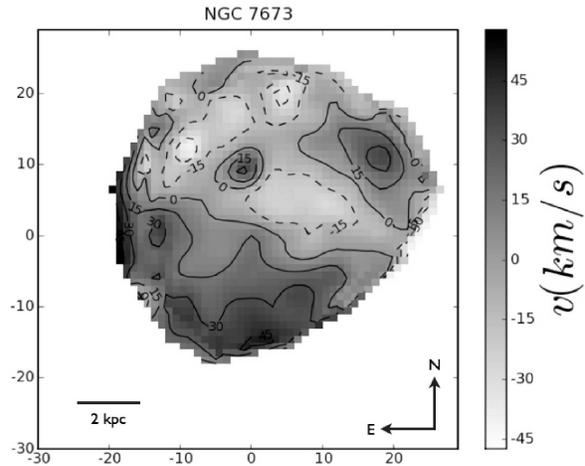}
\caption{PPAK H$\alpha$ velocity map of NGC 7673 using the low resolution V300 configuration. The velocity map shows an asymmetric velocity field with at least two independent kinematic components in the northern side of the galaxy. \label{fig:velmap}}
\end{center}
\end{figure}

The velocity map of NGC 7673 can be seen in Figure~\ref{fig:velmap}. Measurements of the velocity down to our S/N limit were possible for a region extending 40 arcsec in diameter. This compares well with the effective radius of the galaxy, which is 8 arcsec ($r_e=1.9$~kpc). Further away, the S/N was not high enough to proceed to the measurement of the velocity. Border and pixelization effects due to the interpolation process were not considered in the analysis. These effects manifest as either a steep increase or steep decrease of the velocity towards the edges of the available data. 

The velocity map shows an asymmetric velocity field that ranges from approximately -45 to 45~km~s$^{-1}$. The velocities shown in the map are referenced to the redshift of the central peak velocity width of the galaxy (see Section~\ref{kinvelwid}). No absolute calibration of the recession velocity was attempted. This redshift is 0.011293, smaller, but within two sigma of the one available in the Nasa Extragalactic Database ($z=0.011368$).

\begin{figure}
\begin{center}
\includegraphics[width=8cm]{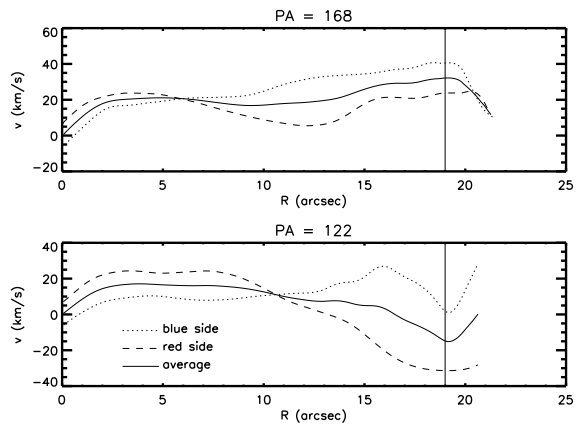}
\caption{Rotation curves along the major axis of NGC 7673 for two different position angles. From our data we derive a PA equal to 168$^{\circ}$, versus $122^{\circ}$ from the Hyperleda catalog. The blue side (\textit{dashed}), red side (\textit{dotted}), and an average of both sides (\textit{solid}) of the galaxy are plotted. \label{fig:rotcurv}}
\end{center}
\end{figure}

The existence of an almost straight velocity contour that crosses the galaxy from East to West through the central velocity width peak suggests that it actually traces the position of the minor axis of the galaxy, yielding a major axis position angle (PA) equal to 168$^{\circ}$. A PA of 122$^{\circ}$ derived from optical images (taken from HyperLeda\footnote{http://leda.univ-lyon1.fr/}) was used in past studies of this galaxy. A rotation curve can be derived for both position angles by reading the measured velocities along the corresponding major axis (see Figure~\ref{fig:rotcurv}). A PA equal to 168$^{\circ}$ provides a smoother rotation curve from which an uncorrected by inclination rotation velocity of about 30~km~s$^{-1}$ can be inferred.

While the southern half of the galaxy is consistently moving away from the observer, the northern half is not uniformly moving towards. Two discrete regions located in the northern half of the galaxy are moving away from the observer in opposite direction to their surroundings. 

First, we measure a smaller circular region, about 1~kpc wide, about 10~arcsec north of the centre of the FOV. This region is moving away from the observer at about 35~km~s$^{-1}$ with respect to its surroundings. Furthermore, this region, although confined to only the area of one fiber, was detected on the three independent dithering exposures. Second, we identify a bigger elongated region, about 3.5~kpc long, which is located towards the northwest, about 2 kpc away from clump F. This is too far from the edge of the available data for us to consider it as a border effect. This region, considerably more extended than the previous one shows a smoother velocity gradient. Nevertheless, it also peaks at around 35~km~s$^{-1}$  with respect to its surroundings. We discuss below whether these kinematically decoupled components show physical properties that differ from the rest of the galaxy.

\begin{figure}
\begin{center}
\includegraphics[width=8cm]{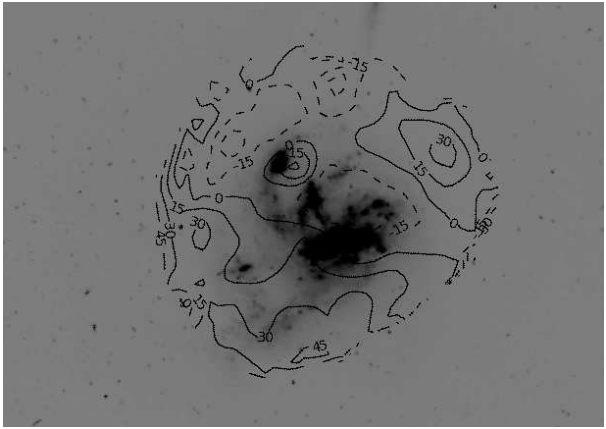}
\caption{PPAK H$\alpha$ velocity map contours overlaid on the F555W WFPC2 image of NGC 7673. Clump B from \citet{duflot82} can be identified with one of the independent kinematic component. There is no optical counterpart of the second one. \label{fig:velhst}}
\end{center}
\end{figure}

When compared to the F555W WFPC2 image of NGC 7673 (see Figure~\ref{fig:velhst}) two characteristics stand out: (i) the velocity field is considerably more extended than the region where the more luminous star forming regions are confined; and (ii) while the smaller decoupled region appears to be linked to clump B, no optical counterpart can be linked to the bigger one.

\begin{figure}
\begin{center}
\includegraphics[width=8cm]{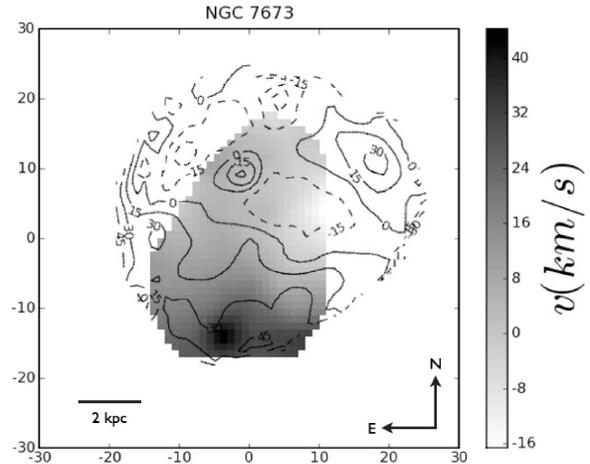}
\caption{PPAK H$\alpha$ velocity map contours of NGC 7673 overlaid on the DENSEPAK H$\alpha$ velocity map of NGC 7673 from HG99. Both velocity maps are consistent within their respective uncertainties. \label{fig:dpak}}
\end{center}
\end{figure}

We compare our results to those of HG99 (see Figure~\ref{fig:dpak}). Using the PPAK H${\alpha}$ image of the galaxy and Figure 3 from HG99, our and their FOV were registered by finding the position of two fibers mapping the same region of the galaxy. The DENSEPAK fiber array was drawn within the rectangular FOV on their figure for that purpose. This was needed because HG99 did not state the pointing position for their observations. A velocity map using their measurements was then produced down to a resolution of 1~arcsec using our interpolation method and reference system. This allowed us to directly compare both maps. The residual image reveals an offset of 20~km~s$^{-1}$ and a standard deviation of 12~km~s$^{-1}$. While their better spectral resolution, as stated above, allowed them to look for multiple kinematic components, both our FOV and spatial sampling are better, which allows us to extend the study of the galaxy. The only notable difference between both maps is the presence of the small circular region in our data at the position of clump B. This could be explained by both the sampling an the dithering technique we used. Given the size of this feature, HG99 might have simple missed it in between their fibers.

\subsection{Neutral Hydrogen Gas} \label{nhg}

\begin{figure}
\begin{center}
\includegraphics[width=8cm]{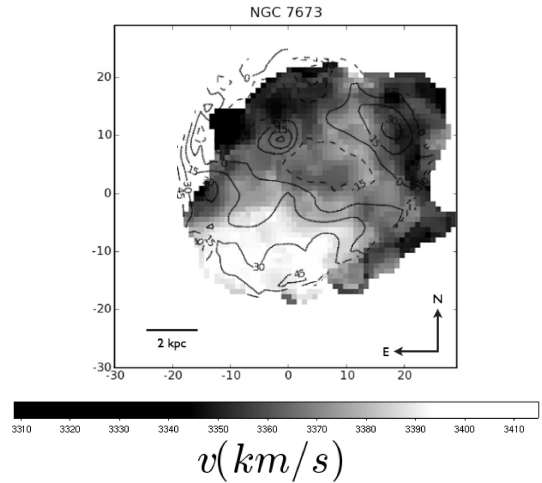}
\caption{PPAK H$\alpha$ velocity map contours overlaid on VLA neutral atomic hydrogen velocity map of NGC 7673. The combined B+C+D arrays were used for the VLA map. The size of the beam is 6". Both velocity maps are consistent within their respective uncertainties. The offset between both might be explained by the two plane-parallel disks as discussed in the text.  \label{fig:velmapvla}}
\end{center}
\end{figure}

As discussed in Section~\ref{kinvel}, the velocity field of the galaxy derived from the centroids of the H${\alpha}$ emission line profiles shows asymmetries. The neutral hydrogen velocity field resembles the ionized one (see Figure~\ref{fig:velmapvla}) in extent down to the detection limits, overall appearance, and velocity range ($\sim90$~km~s$^{-1}$) (Pisano et al. \textit{in prep.}). 

Both maps were registered using the header coordinates. For the registration process several images were used: F555 HST, H${\alpha}$ PPAK, 6cm VLA, 20 cm VLA, and DSS. The uncertainty in the registration was established to be no higher than half an arcsecond. When the distributions of H {\small{I}} and H {\small{II}} are compared there is an obvious slight offset between them. The H {\small{II}} distribution extends slightly more towards the northeast while the H {\small{I}} one extends considerably more towards the west. This difference might be explained by two plane-parallel disks. The H {\small{II}} disk might be thicker than the H {\small{I}} and only the nearest side might be visible to us. If an inclination of 45$^{\circ}$ is assumed for both disks, a separation of about 1.7~kpc can be derived between the nearest side of the thicker H {\small{II}} disk and the H {\small{I}} disk. This is in agreement with the typical scale height of a thick disk \citep{howk00}. Nevertheless, the uncertainties associated with these estimations are quite large.  

On top of that, a residual image was generated by subtracting the optical velocity map from the radio one for the shared area. The resulting mean and standard deviation are 0 and 15~km~s$^{-1}$, which make them consistent with each other. When only the receding part of the galaxy was considered these numbers were 8 and 9~km~s$^{-1}$ respectively indicating that there is not a cancellation effect between the receding and the approaching sides of the galaxy.

On the other hand, while the distribution of the H {\small{I}} resembles that of the ionized gas, the H$_2$ distribution, as traced by CO observations, is concentrated along clump A \citep{garland05}; no CO was detected in clump B. It is to be noted that the higher column densities of H {\small{I}} trace the location of CO. This lack of CO detection in clump B could be explained if the burst was quenching, opaque to CO radiation, or being injected with H {\small{I}} through galactic winds. At the current SFR (CM09) the H {\small{I}} still present in LCBG NGC7673 ($M_{\mathrm{H {\small{I}}}}=4.09 \times 10^{9}$~M$_{\odot}$ according to \citet{pisano01}) would be exhausted in about 1~Gyr. 

\subsection{Velocity Width Map} \label{kinvelwid}

\begin{figure}
\begin{center}
\includegraphics[width=8cm]{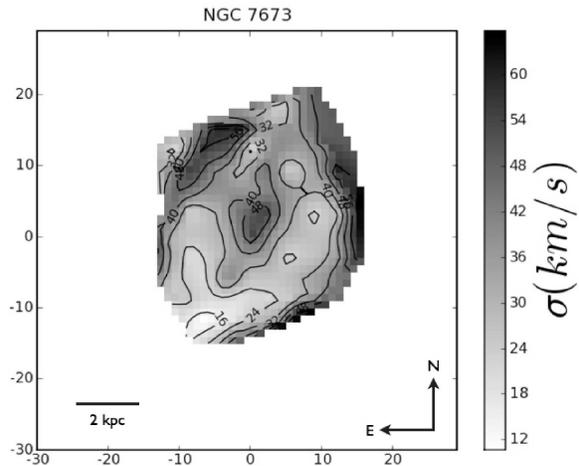}
\caption{PPAK [OIII]$\lambda$5007 velocity width map of NGC 7673 using the V1200 configuration. The central peak roughly coincides with the luminosity-weighted center (photometric) and the center of the outer-isophotes  (geometric) of the galaxy, and is assumed to be the dynamical center of the galaxy. Furthermore, the minor axis of the galaxy as derived from the velocity contours goes through this peak. \label{fig:sigmap}}
\end{center}
\end{figure}

\begin{figure}
\begin{center}
\includegraphics[width=8cm]{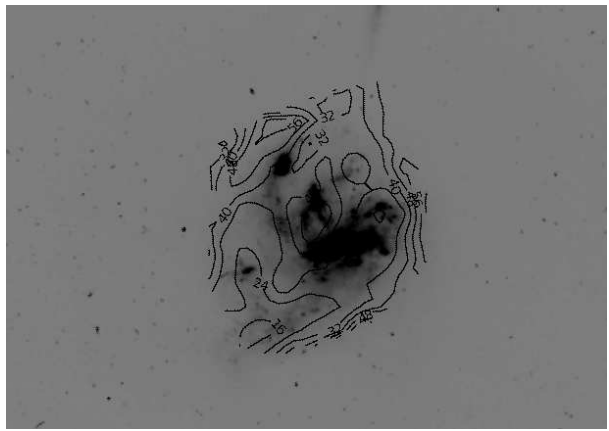}
\caption{PPAK [OIII]$\lambda$5007 velocity width map contours overlaid on the F555W WFPC2 image of NGC 7673. North is to the top and East is to the left. \label{fig:sighst}}
\end{center}
\end{figure}

The velocity width map of the galaxy (see Figure~\ref{fig:sigmap} and Figure~\ref{fig:sighst}) can be used in combination with the velocity map to better understand the kinematic properties of the galaxy. In virialized systems the velocity width peak tends to coincide with the centroid of the velocity map, assuming it is organized rotation which dominates the galaxy motions. 

The velocity width map of NGC 7673 peaks ($\sigma=54\pm6$~km~s$^{-1}$) around the centre of the FOV (see Figure~\ref{fig:sighst}), between clumps A and C. This position roughly coincides with the luminosity-weighted centre (photometric) and the centre of the outer-isophotes  (geometric) of the galaxy. These are located respectively about 3 and 4~arcsec (920 and 690~pc respectively) towards the west of the centre of the FOV; these offsets are comparable to the size of one fiber. The geometric centre was found as the average of the position of each fiber with signal from the galaxy; while the photometric centre was found as the average position of each fiber with signal from the galaxy weighted by its integrated flux. Two additional elongated peaks in velocity width as large as the central one can be found to the northeast and northwest of the galaxy. The velocity width of the galaxy ranges from about 15 to 60~km~s$^{-1}$ within a FOV covered by fibers for which S/N is at least 30. 

As for the integrated velocity width HG99 found $W_{20}=126$~km~s$^{-1}$ ($v_{rot}=0.5W_{20}/\sin{i}$). This is slightly larger than the value found by \citet{pisano01}, $W_{20}=119$~km~s$^{-1}$ from Keck echelle spectroscopy. Furthermore, \citet{pisano01} find the H {\small{II}} profile to be narrower than the H {\small{I}} profile (119 versus 164~km~s$^{-1}$). From our integrated velocity width measurement, we find $W_{20}=159$~km~s$^{-1}$, closer to \citet{pisano01}. We estimate $W_{20}$ by measuring $\sigma$ and assuming a gaussian profile ($W_{20}=2\sqrt{2\ln{5}}\times\sigma$). Differences with HG99 might be attributed to our larger FOV, which, on the other hand, is large enough to resemble that of the H {\small{I}} observations of \citet{pisano01}. The integrated velocity width of the galaxy was also measured after removing clump B without recording meaningful differences in our measurement.

\section{Discussion}

\subsection{Minor Merger Scenario}\label{mms}

As stated above, two independent kinematic components are found in NGC 7673: a compact one with an optical counterpart (clump B); and an elongated one without one. The independent kinematic component found at the location of clump B is interesting not only because of its optical counterpart but because of further evidence for decoupling that we discuss below. 

A first attempt to investigate the nature of clump B was made by calculating the [OIII]$\lambda$5007/H$\beta$ and [NII]/H$\alpha$ ratios for the entire galaxy, and this component in particular, to study the possible presence of AGN activity. Nevertheless, these ratios, around 0.3 and -0.7~dex, respectively, are in agreement with those of an H {\small{II}} region \citep{osterbrock89}. Furthermore, the pressure derived for clump B from the [SII]/H$\alpha$ ($\sim0.4$) and [NII]/H$\alpha$ ($\sim0.2$) ratios is also in agreement with that of a starburst \citep{rickes08}. For clump B and the entire galaxy those ratios are considerably smaller than those found in LINERs, in agreement with those found in starburst galaxies, and only close to those found in Seyfert galaxies (see Figure~12 in \citet{rickes08}).   

As mentioned above, HG99 were able to fit the H$\alpha$ emission lines of their IFU data with two Gaussian profiles throughout most of the galaxy. In particular, before HG99, \citet{taniguchi87} found two (one narrow, one broad) kinematic components at the location of clump B, in the H$\alpha$ emission line. Nevertheless, after simulating those by using their measurements in combination with our [OIII]$\lambda$5007 emission lines, spectral resolution, and noise, only a marginal detection would have been possible. Thus, the possibility of finding a second broad component at the position of clump B by means of the analysis of the profiles of our emission lines was rejected.

Another possibility is that clump B peculiar kinematics are due to galactic winds and even though, as mentioned above, the quality of our data keeps us from establishing any final word about the nature of this region, a further qualitative analysis is possible based on the H$\beta$ luminosity ($L_{\mathrm{H}\beta}$) measured by CM09 using

\begin{math}
Q=\frac{L_{\mathrm{H}\beta}}{h\nu_{\mathrm{H}\beta}}\frac{\alpha^{eff}_{\mathrm{H}\beta}}{\alpha_B}
\end{math}    

\citep{osterbrock89}, where $Q$ is the number of photons harder than Ly${\alpha}$ emitted by a star forming region, $h$ is the Planck constant, $\nu_{\mathrm{H}\beta}$ is the frequency of H$\beta$, $\alpha^{eff}_{\mathrm{H}\beta}$ is the case B H {\small{I}} recombination coefficient for H$\beta$, and $\alpha_{B}$ is the recombination coefficient for H-like ions. Assuming then that all the stars in the star-forming region are O7 with a mass of 60 $M_{\odot}$ and a Salpeter Initial Mass Function \citep{salpeter55} we find that 18500 stars are responsible for the ionization of the gas. While this is an approximation the presence of Wolf-Rayet features in our spectra (CM09) is consistent with massive stars. If all those stars were to explode as supernovae with an energy of $10^{51}$~erg, the total thermal energy released ($E=1.85\times10^{55}$~erg),  would be unequivocally smaller than the binding energy ($\Omega_{G}=5.23\times10^{57}$~erg) as defined by \citep{yoshii87}. Galactic winds cannot be then responsible for the kinematic properties of the component associated with clump B.  If we assume the total thermal energy released is transformed into kinematic energy, velocities up to 20~km~s$^{-1}$ are accounted for. Such a low value is in disagreement with the broad component of \citet{taniguchi87} and could not be detected with our spectral resolution.

On the other hand, as stated by HG99, both photometric and spectroscopic similarities between NGC 7673 and NGC 3310 (a well known system largely classified as a minor merger) in both H {\small{II}} and {H \small{I}} data leads to consider NGC 7673 as a candidate for a major starburst triggered by a minor merger. This minor merger with a dwarf companion would account for the different bursts in the inner parts of the galaxy and the different morphological features, such as arcs, in its outer parts. 

\begin{figure}
\begin{center}
\includegraphics[width=8cm]{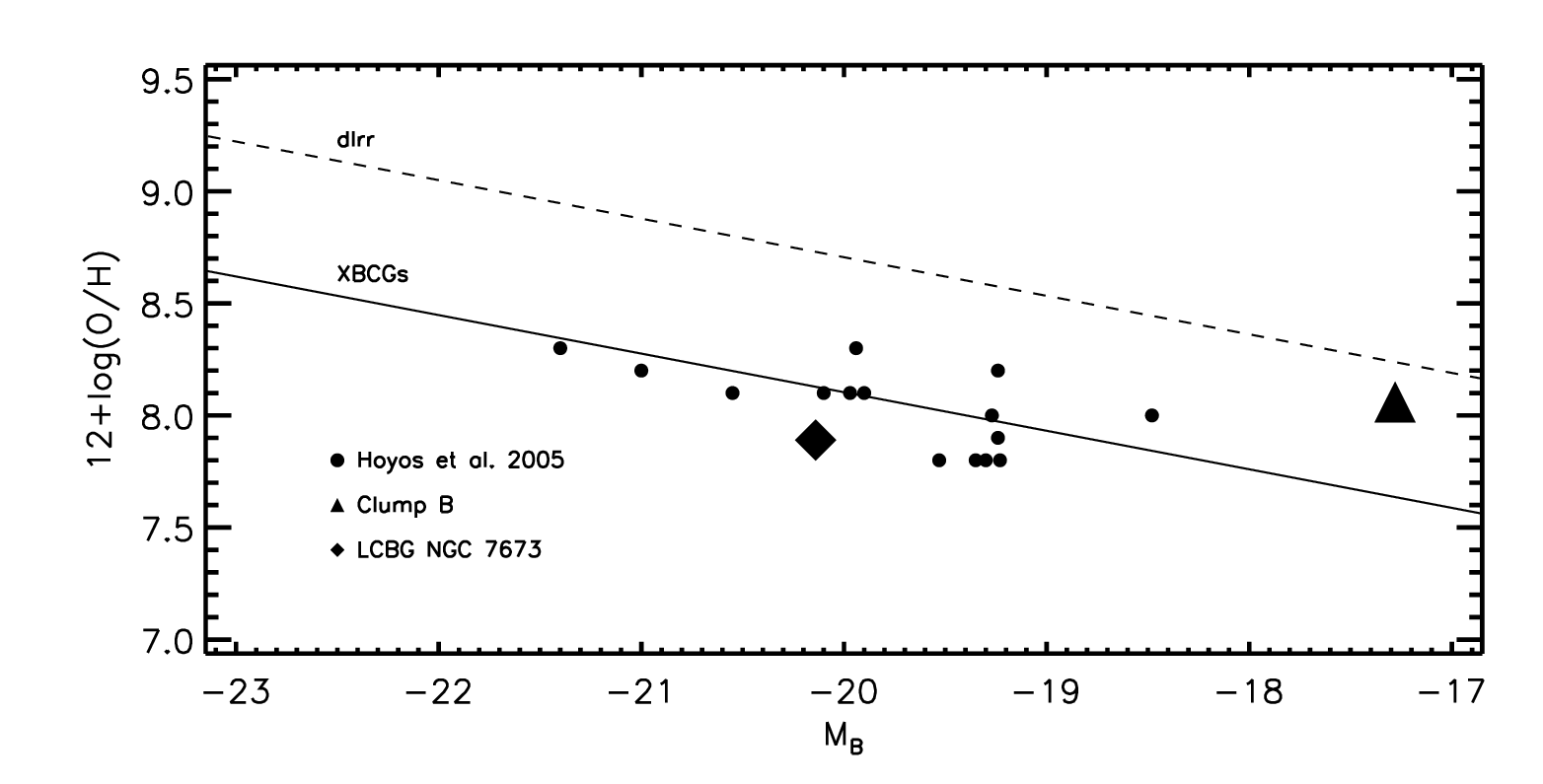}
\caption{12 + log (O/H) vs. $M_B$ for a sample of intermediate-redshift LCBGs \citep{hoyos05}. The position of NGC 7673 and clump B are also shown as measured by CM09. The dashed line represents the luminosity-metallicity relation for local dIrr \citep{richer95}. The solid line represents this relation for extremely metal-poor BCGs \citep{kunth00}. The uncertainties of our metallicity and luminosity measurements are smaller than our symbols. \label{fig:metvslum}}
\end{center}
\end{figure}

\begin{figure}
\begin{center}
\includegraphics[width=8cm]{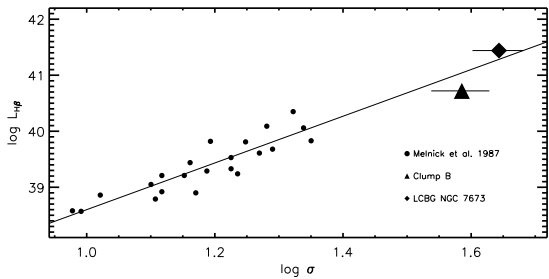}
\caption{$L_{H{\beta}}$ vs. $\sigma$ for a sample of giant HII regions \citep{melnick87}. The position of NGC 7673 and clump B are also shown as measured by this work ($\sigma$) and CM09 ($L_{H{\beta}}$). The solid line represents the correlation between these parameters. This correlation holds from giant HII regions to similar starburst galaxies observed at high redshift \citep{siegel05}. The uncertainties of our luminosity measurements are smaller than our symbols. \label{fig:lumvssig}}
\end{center}
\end{figure}

Metallicity and continuum (5600 \--- 5800~{\AA}) measurements of CM09 show marginally higher metallicity values, and a secondary peak (second in intensity after the central one) at the location of clump B respectively. The metallicity values found by CM09 for clump B and its surroundings from [OIII]$\lambda$4363 are respectively $8.07\pm0.06$ dex and $7.76\pm0.06$~dex. These could be explained by the presence of an extremely giant H {\small{II}} region at the location of clump B within a lower metallicity environment. When we compare the metallicity and $M_B$ of clump B with a sample of intermediate-redshift LCBGs from \citet{hoyos05} (see Figure~\ref{fig:metvslum}) clump B, less luminous and slightly less metallic, falls closer to dwarf irregulars while NGC 7673 falls within the region occupied by intermediate-redshift LCBGs, 

The H$\beta$ luminosity and the velocity width of clump B follow the correlation found by \citet{melnick87} for nearby giant H {\small{II}} regions (see Figure~\ref{fig:lumvssig}). Nevertheless, clump B is brighter and considerably more massive than any of the nearby giant H {\small{II}} regions within their sample. Knowing that this correlation holds from nearby H {\small{II}} regions and galaxies to the H {\small{II}}-like galaxies found at intermediate-, and high-redshift \citep{siegel05}, suggests that an infalling dwarf galaxy, instead of a giant H {\small{II}} region at the location of clump B might be responsible for its peculiar behavior. If the minor merger scenario were to be confirmed, the elongated independent kinematic component cannot be disregarded as a possible side effect within the minor merger scenario.  

Furthermore, CM09 measurements of the equivalent widths of the underlying population absorption spectral features (e.g., H$\beta$, H$\gamma$) show a peculiarity at the location of clump B. While the strength of these features is noticeable throughout the galaxy, it is almost nonexistent at the location of clump B. The strength of the equivalent width of these features account for the age of the underlying population. Strong lines are mainly due to a population of A class stars and correspond to systems a few hundred thousand years old. Weaker lines are typical of either younger or older populations. Nonetheless, clump B shows signs of decoupling or differentiation, at the least marginally, with respect to its environment from the point of view of both the gas and the stars. Notice that as stated above no optical counterpart can be linked to the bigger elongated region, and we found no signs of decoupling or differentiation other than the kinematic signature.

\subsection{Mass}\label{mass}

Galactic kinematics play an important role on the measurement of masses of nearby and distant galaxies. Galaxies' masses are related to their kinematics via the Virial Theorem. Nevertheless, we need to be careful when making the assumption that an object is virialized. By studying LCBG NGC7673, among other examples of local counterparts of the distant starburst population, we can learn whether or not most of these objects are virialized and how careful we need to be when inferring their masses from their kinematics.   

If we assume for NGC 7673 a peak to peak velocity range as suggested by the rotation curve derived above ($\sim60$~km~s$^{-1}$), the inferred rotational velocity $v_{rot}=\sqrt{\frac{M_{dyn}G}{R_e}}$ translates into a dynamical mass within $R_e$ ($R_e=1.9$~kpc) of around $8.0\times10^{8}~M_{\odot}$. \citet{pisano01} inferred a dynamical mass of $2.5\times10^{10}~M_{\odot}$  within $R_{\mathrm{H {\small{I}}}}$ ($R_{\mathrm{H {\small{I}}}}=8.3$~kpc $\sim4.5\times R_e$), or $5.5\times10^{9}~M_{\odot}$  within $R_e$. In both cases an inclination of 45$^{\circ}$ is assumed. 

On the other hand, considering the K magnitude of NGC 7673 and assuming the mass luminosity relation $M/L_K=0.51$ for BCDs from \citet{perez03} we derive a stellar mass of $1.5\times10^{10}~M_{\odot}$. A total mass of  $1.9\times10^{10}~M_{\odot}$ is then calculated by adding the mass of the neutral hydrogen estimated by \citet{pisano01} ($M_{\mathrm{H {\small{I}}}}=4.09\times10^{9}$). If the system is to be virialized the resulting rotational velocity would be $v_{rot}=133$~km~s$^{-1}$. In order for the rotation curve derived for a PA equal to 168$^{\circ}$ to agree with this $v_{rot} $ the inclination of the galaxy must be close to 15$^{\circ}$, which on the other hand is not unreasonable when one takes a look at the optical image of the galaxy. Nevertheless, this comparison depends strongly in a well established mass luminosity relation and current studies (Matthew Bershady, private communication) seem to indicate that the mass luminosity relation for this particular kind of galaxies might be as much as five times lower, which would match better our photometric and dynamical mass estimates.

From the velocity width measurements of our integrated PPAK data we find $W_{20}=159\pm4$~km~s$^{-1}$. If an inclination equal to 15$^{\circ}$ is also to be considered 
and we use the relation between $W_{20}$ and $v_{rot}$ used by \citet{pisano01}, 
the rotational velocity would rise to $v_{rot}=307$~km~s$^{-1}$. 
On the other hand, a recent study by \citet{andersen09} show that for low-inclination systems like NGC 7673, $2.9~v_{rot}=W_{20}/\sin{i}$. This provides an uncorrected for inclination $v_{rot}=55~km~s^{-1}$, which is closer to the uncorrected value we derived from our rotation curve ($30~km~s^{-1}$).  

As it is, neither the peak to peak velocity range shown by the optical data nor the one shown by the radio data, seems to account for such a large $v_{rot}$ even though the entire FOV of NCG 7673 is covered in both cases. This suggests that $W_{20}$ and, therefore, FWHM and $\sigma$, might not good indicators of the rotational velocities and subsequent dynamical mass of starburst galaxies like NGC 7673, and mostly accounts for the presence of kinematically decoupled components and asymmetries alike, and their position within the galaxy.






\section{Conclusions}

From our study we conclude that even though NGC 7673 shows an asymmetrical velocity map, a dynamical centre of the galaxy and a position angle can still be inferred. The former coincides roughly with the geometrical and photometrical ones, while the latter is found to be 168$^{\circ}$. Using this centre if we consider a rotation curve along the major axis of the galaxy we derive a rotational velocity of about 30~km~s$^{-1}$ before inclination corrections. For this rotational velocity to account for the total mass of the system inferred from the K band magnitude and H{\small{I}} observations an inclination of $15^{\circ}$ is necessary. 

The LCBG population may or may not consistently show asymmetrical velocity fields, but being NGC 7673 a prototypical LCBG and taking into account recent results by \citet{ostlin04} and \citet{cumming08} on two LCBGs, and preliminary results on other objects in our sample, we are inclined to believe that this might be a trend among this population of star forming galaxies. This behavior turns our data set into essential when approaching the study of these objects and showing the importance this effect might have when trying to derive rotational velocities and dynamical masses from long-slit spectroscopy of distant vigorous starburst galaxies, including LBGs. Our data set has both a wide FOV to detect the galaxy down to the low surface brightness outskirts, and a high spatial resolution that allows for small decoupled kinematic components to be found. An example is the compact small decoupled kinematic component found at the location of clump B in NGC 7673. As discussed in Section~\ref{mms}, clump B shows no evidence for neither AGN activity nor SNe galactic winds and is in agreement with being either an extremely giant H {\small{II}} region or an infalling dwarf galaxy. Even though the latter scenario is plausible both spectroscopically and photometrically, a high spectral resolution study is still necessary in to confirm it.

\section*{Acknowledgments}

It is a pleasure to thank the many people who welcomed our project into the 3.5-m telescope in CAHA, where we always felt at home, especially those cozy snowy nights full of pastries and hot chocolates. We would especially like to thank Ana Guijarro, Jes\'us Aceituno, and Santos Pedraz. We also thank Carlos Hoyos for interesting and highlighting discussions. Finally, we thank our referee, Matthew Bershady, for his helpful comments and suggestions. J. P\'erez-Gallego acknowledges support from a University of Florida Alumni Fellowship and R. Guzm\'an from NASA Grant LTSA NA65-11635. This work is partially funded by the Spanish MICINN under the Consolider-Ingenio 2010 Program grant CSD2006-00070: First Science with the GTC (http://www.iac.es/consolider-ingenio-gtc). This work is partially funded by the Spanish Programa de Astronom\'ia y Astrof\'isica under grants AYA2006-02358 and AYA2006-15698-C02-02.

\label{lastpage}

\end{document}